\documentclass[num-refs]{wiley-article}
\usepackage{siunitx}
\usepackage{amsmath,amssymb}
\usepackage{verbatim}
\usepackage{subcaption}
\usepackage[section]{placeins}

\papertype{Original Article}
\paperfield{Journal Section}

\title{Recurrent Neural Network-based Model Predictive Control for Continuous Pharmaceutical Manufacturing}

\abbrevs{Active Pharmaceutical Ingredients, APIs; Model Predictive Control, MPC; Critical Quality Attributes, CQAs; Artificial Neural Networks, ANNs; Recurrent Neural Networks, RNNs}

\author[1]{Wee Chin Wong}
\author[1]{Jiali Li}
\author[1]{Xiaonan Wang}

\affil[1]{Department of Chemical \& Biomolecular Engineering, Faculty of Engineering, National University of Singapore. 4 Engineering Drive 4, Singapore 117585, Singapore}
\corraddress{Xiaonan Wang, Department of Chemical \& Biomolecular Engineering, Faculty of Engineering, National University of Singapore. 4 Engineering Drive 4, Singapore 117585, Singapore}
\corremail{chewxia@nus.edu.sg}

\fundinginfo{\textcolor{black}{The authors thank the MOE AcRF Grant in Singapore for financial support to the projects of Precision Healthcare Development, Manufacturing and Supply Chain Optimization (R‐279‐000‐513‐133) and Advanced Process Control and Machine Learning Methods for Enhanced Continuous Manufacturing of Pharmaceutical Products (R-279-000-541-114).}}

\runningauthor{Wong et al.}

\begin{document}

\maketitle

\begin{abstract}
The pharmaceutical industry has witnessed exponential growth in transforming operations towards continuous manufacturing to effectively achieve increased profitability, reduced waste, and extended product range. Model Predictive Control (MPC) can be applied for enabling this vision, in providing superior regulation of critical quality attributes. For MPC, obtaining a workable model is of fundamental importance, especially in the presence of complex reaction kinetics and process dynamics. Whilst physics-based models are desirable, it is not always practical to obtain one effective and fit-for-purpose model. Instead, within industry, data-driven system-identification approaches have been found to be useful and widely deployed in MPC solutions. In this work, we demonstrated the applicability of Recurrent Neural Networks (RNNs) for MPC applications in continuous pharmaceutical manufacturing. We have shown that RNNs are especially well-suited for modeling dynamical systems due to their mathematical structure and satisfactory closed-loop control performance can be yielded for MPC in continuous pharmaceutical manufacturing.

%textit{Pharmaceutical, Critical Quality Attributes, Good Manufacturing Practice, Deep-Learning, Recurrent Neural Networks, System Identification, Model Predictive Control}

% Please include a maximum of seven keywords
\end{abstract}

%%%%%%%%%%%%%%%%%%%%%%%%%%%%%%%%%%%%%
\section{Introduction}
%%%%%%%%%%%%%%%%%%%%%%%%%%%%%%%%%%%%%
The pharmaceutical industry has a growing interest in addressing key challenges such as the reduction of waste, cost, footprint and lead-times by implementing the concept of continuous manufacturing \cite{Lakerveld2015a,Schaber2011}. As a result, with the development in continuous manufacturing technologies and quality by design paradigm, there is a trend of increasing demand for more advanced model identification and process control strategies in the continuous pharmaceutical industry \cite{Lakerveld2015a}. 

One emerging application area for continuous manufacturing is to extend the palette of permissible reaction conditions and enable reaction outcomes that are quite challenging when performed under batch conditions \cite{Glasnov2016,Gutmann2015,Poechlauer2013}. The production of Active Pharmaceutical Ingredients (APIs) is the core part of pharmaceutical manufacturing. As a result, many researchers have proposed  or employed continuous manufacturing technologies in the production processes of complex APIs. The fact is that these reactions are generally both reversible and coupled with many side reactions, thereby leading to a highly non-linear nature of the reaction system. As a result, this imposes difficulty on model identification. In addition to that, the need to operate in tight and extreme conditions leads to needs in effective and sophisticated control strategies \cite{Benyahia2012b,F.Susanne2017,Mascia2013}. The control and model identification of these continuous API reactions are particularly challenging due to the complexity of the underlying phenomena and the associated ultimate impact on reaction yields, molecular structure, downstream processing and Critical Quality Attributes (CQAs) of final product. 

Model-based advanced control technology is viewed to be vital for enabling continuous pharmaceutical manufacturing by improving control of CQAs. The system identification is an indispensable part of the control performance for the whole process. Many rigorous models have been proposed to describe different API reactions \cite{Benyahia2012b,S.Brueggemeier,Mesbah2017a}. These physics-based models have a clear physical interpretation. However, they can suffer from a complicated structure and may incur excessive computational cost during on-line control. Moreover, in many nonlinear systems, it is extremely difficult and expensive to obtain an accurate model of the process from first principles \cite{M.Hussain1999}. In contrast, experimental, data-driven heuristic models are easier to get and with relatively simple mathematical descriptions \citep{L.cheng2015}. Belonging to the data-driven methods, neural networks have been applied actively in chemical and biochemical processing industry, especially has been implementing in some complex non-linear processes where process understanding is limited \citep{Z.Xiong2005,Tian2001,M.a.Hussain2001,Nagy2007a}.

System identification in chemical engineering applications addresses temporal correlations and interactions amongst the various states and controlled variables of concern. It is generally recognized that Artificial Neural Networks (ANNs) have the universal capability of approximating any function. Compared to the typical feed-forward ANNs, the structure of RNNs is such that the latter are better suited for providing temporal predictions. RNNs are inherently structured to use the hidden variables as a memory to capture long-term dependencies and have self-feedback of neurons between and within the hidden layers. This feedback process means RNNs inherently possess a "deep" structure. RNNs have been widely used in sequence learning problems including scene labeling \cite{byeon2015} and language processing \cite{Cho2014}, with good success. The authors of \cite{Lee2017a} provide an excellent overview of the applications of neural networks especially from a process systems engineering perspective. Thus, in our work, we study the implementation of a RNN for model identification, using a complex reaction in a single Continuous-Stirred Tank Reactor (CSTR) for pharmaceutical API production, as an exemplary reference.

%%This satisfactory result shows the potential of applying RNN into tackle more complex nonlinear biological reaction systems that are increasingly applied in development of more efficient production pathway of some complex APIs, production of macromolecular drugs as well as in advanced biotechnology applications \cite{Paddon2014,June2018,N.Baeshen2014}.

Apart from the modeling of complex pharmaceutical reactions, how to design effective model-based control algorithms for continuous manufacturing is also a research focus in academic literature \cite{Rehrl2016} and industry. In addition, the United States Food and Drug Administration (FDA) has emphasized that, in order to take full advantage of the real-time release Process Analytical Technology (PAT) concept, continuous-flow reactions require on-line monitoring and feedback control \cite{FDA2004}. Model-based control is an excellent framework to rigorously incorporate the information of new and existing PAT technologies, as they emerge or get developed.

The MPC method is widely used in industrial applications \cite{J.B.Rawlings2009,P.Tatjewski2007,GARCIA1986} and it has been applied in control of CQAs in continuous pharmaceutical manufacturing \cite{Mesbah2017a,Rehrl2016}. In \cite{Mesbah2017a}, the authors presented two plant-wide MPC designs for whole continuous pharmaceutical manufacturing pilot plant from chemical synthesis to tablets formation using the quadratic dynamic matrix control algorithm, an earlier form of MPC. It shows by monitoring CQAs in real time, Critical Process Parameters (CPPs) can be actively manipulated to enable more robust and flexible process operation via feedback control. The advantages of MPC compared to conventional Proportional-Integral-Derivative (PID) controllers are highlighted in \cite{Rehrl2016} via comparison of the control strategies to a feeding blending unit used in continuous pharmaceutical manufacturing. In addition to the work done on MPC, RNNs had been proposed to be used in conjunction with MPC in previous studies. For example, \cite{pan2012} investigated the applicability of RNNs for control of non-linear dynamical systems. RNNs have been used for predictive control of CSTRs (for instance \cite{Seyab2008}) but these have been limited to relatively basic kinetics (e.g., single-step, irreversible kinetics $A \rightarrow B$). Furthermore, these MPC studies have rarely been applied to the most challenging and essential part in continuous pharmaceutical manufacturing of reactor control. %In \cite{Jeong2001}  and \cite{Nagy2007} MPC has been applied in a continuous reactor and a batch process, however the reaction is more about polymerization which is not a typical reaction type for pharmaceutical manufacturing.

Our contribution is in demonstrating the applicability of a special class of neural networks, termed RNNs, for MPC applications relevant to relatively complex continuous pharmaceutical manufacturing reaction applications. This paper is organized as follows. Section \ref{sec:PlantModelNStudies} describes the plant model for system identification and case studies for control closed loop performance assessment. Section \ref{subsec:SysID} explains the RNN-based system identification methodology. Section \ref{sec:MPCformulation} contains the MPC formulation. Results and discussions are shown in Section \ref{sec:resultsandDiscussion}.
% * <xnwang@ucdavis.edu> 2018-05-13T07:58:49.540Z:
% 
% > Our contribution is in demonstrating the applicability of a special class of Neural Networks (NNs), termed Recurrent Neural Networks (RNNs), for MPC applications relevant to relatively complex continuous pharmaceutical manufacturing reaction applications.
% It is good to add a paragraph about our contributions. However, the current writing is abrupt and too brief.  RNNs have already been introduced in details.  Consider to expand here how we fit the research gap. Also, I changed the mis-use of acronyms and space directly in many places.
% 
% ^ <weechin.wong@gmail.com> 2018-05-17T08:39:09.953Z:
% 
% Done.
%
% ^ <weechin.wong@gmail.com> 2018-05-17T08:39:11.465Z.

%RNNs had been proposed to  be used in conjunction with MPC. For example, \cite{pan2012} investigated the applicability of RNNs for control of non-linear dynamical systems. RNNs have been used for predictive control of CSTRs (for instance \cite{Seyab2008}) but these have been limited to relatively basic kinetics (e.g., single-step, irreversible kinetics $A \rightarrow B$). 
 
%%%%%%%%%%%%%%%%%%%%%%%%%%%%%%%%%%%%%%%%%%%%%%%%%%
\section{Plant Model and Control Case Studies} \label{sec:PlantModelNStudies}
%%%%%%%%%%%%%%%%%%%%%%%%%%%%%%%%%%%%%%%%%%%%%%%%%%

The following subsections describe (i) the true plant model for the purpose of simulation and system identification and (ii) case studies in the case of closed-loop control.

\subsection{Plant Simulator} \label{subsec:plantmodel}
We consider the following reaction system in the context of a single CSTR, as was previously studied by \cite{Koppel1982,Seki2004}. Distinct from the literature, our work focuses on system identification using RNNs with a view towards closed-loop control via MPC. Referring to Eq.\ref{eq:rxn}, the feed species is $A$, the desired product is the intermediate $R$ with species $S$ being an undesired byproduct. The reactions are reversible and depending on the value of the manipulated variable(s).%the steady-state concentrations of the various species are very different.
\begin{equation}
A \mathrel{\mathop{\rightleftarrows}^{\mathrm{k_1}}_{\mathrm{k_4}}} R \mathrel{\mathop{\rightleftarrows}^{\mathrm{k_2}}_{\mathrm{k_3}}} S \label{eq:rxn} %%
\end{equation}

The dynamic equations, based on normalized, dimensionless quantities are presented as such:
%% EQN
\begin{eqnarray}
\frac{dC_A}{dt} &=& q[C_{A0}-C_A] - k_1C_A + k_4C_R \nonumber \\
\frac{dC_R}{dt} &=& q[1-C_{A0}-C_R] + k_1C_A + k_3[1-C_A -C_R] - [k_2+ k_4]C_R  \nonumber \\
k_j  &=&   k_{0,j} \exp \bigg\{ \bigg[-\frac{E}{RT_0}\bigg]_j \bigg[\frac{1}{T}-1\bigg] \bigg\},  j\in{1,2,3,4} \nonumber \label{eq:true_plant_ode} \\
\end{eqnarray}
Here, $C_j,  j\in \{A,R,S\}$ refers to the concentration of the respective species within the reactor and form the state vector. The reactions are first-order, reversible with Arrhenius-type temperature dependencies. It is assumed that all concentrations are measured. The manipulated variables (MVs) are feed flow rate ($q$) and reactor temperature ($T$), whose values are also available. Here, it is assumed that the only species entering the CSTR are $A$ and $R$ such that their concentrations sum up to unity. As such, the concentration of species $S$ is easily computed as a function of time. The values of the normalized feed concentrations, Arrhenius pre-exponentials and activation energies can be found in \cite{Seki2004} and are reproduced in the Appendix for ease of reference. 

As shown in Fig. \ref{fig:steadystate}, for a given flow rate, the concentration of the desired product, $C_R$ reaches a maximum at a certain temperature, beyond which  the reaction is driven all the way back to the left of Eq.\ref{eq:rxn}, yielding only feed reactant and no desired product, $R$. This problem possesses rich and relevant dynamics for continuous pharmaceutical manufacturing and is therefore worth detailed investigation. The control problem is challenging due to the existence of input multiplicities \cite{Koppel1982}, \cite{Bequette2007}.
\begin{figure}[bt]
\centering
 \includegraphics[width=\linewidth]{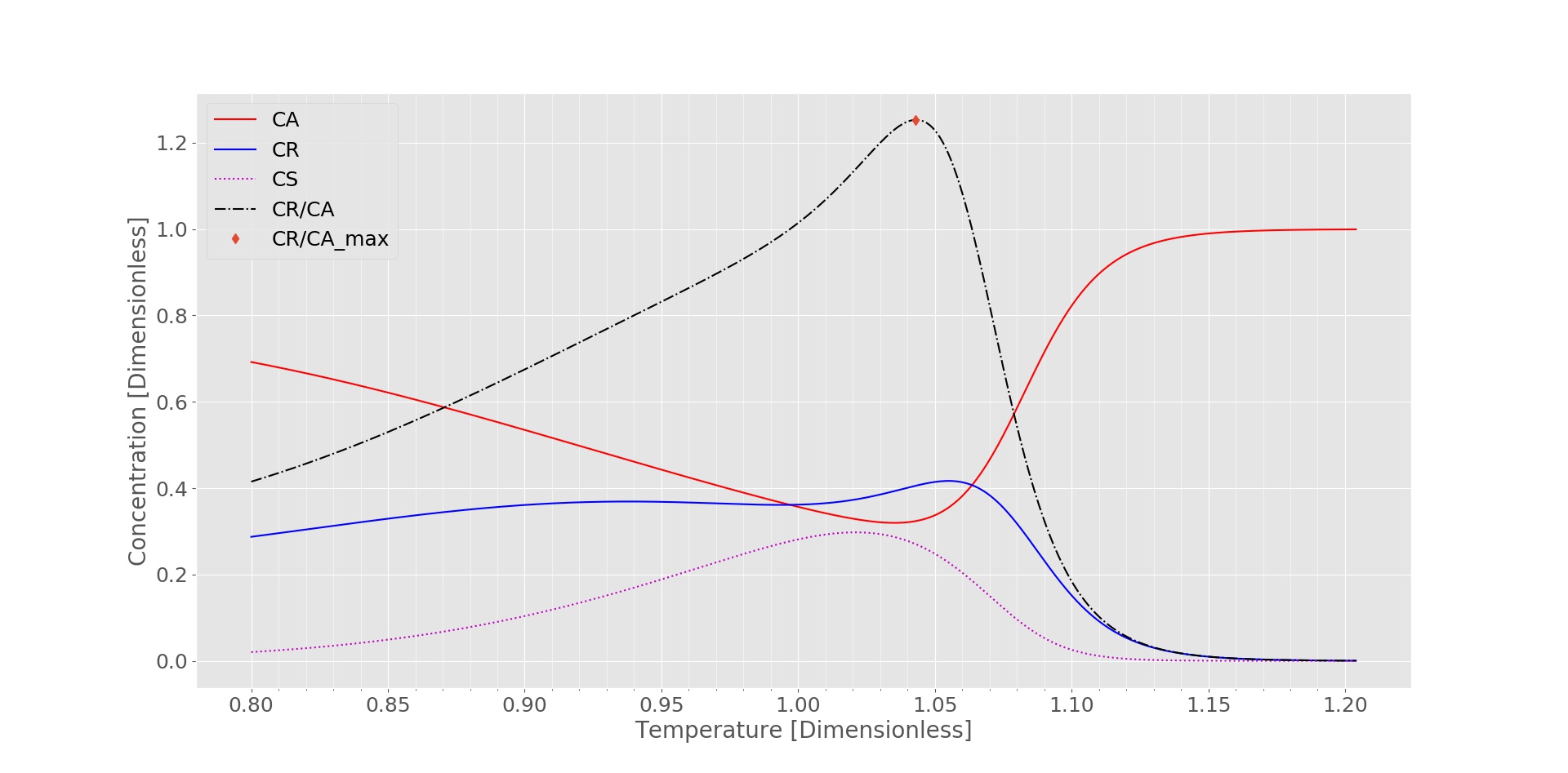} %fig%
\caption{Steady-state conditions as a function of system temperature (with feed flow rate, $q = 0.8$) \textbf{}} \label{fig:steadystate}
\end{figure}
In the sequel, we work in the discrete-time domain with time steps denoted by $k \in \mathbb{Z}^+$, as is common in digital control.  Furthermore, we make the following definitions,  $x \triangleq [C_A,C_R]'$, $u \triangleq [q,T]'$. The underlying plant may be described by a non-linear discrete-time difference equation of the form:
%% EQN
\begin{eqnarray}
x_{k+1} &=& \Phi(x_k,u_k) \nonumber \\
y_k &=& x_k\label{eq:true_plant}
\end{eqnarray}
where $k \in \{0,1,2,\ldots,\}$ is the discrete-time index, $x_k$ the state variables, $u_k$, the manipulated variables. Full state feedback is assumed such that the measured output $y_k$ equals the state vector at all time steps. $\Phi$(.) is the state-transition equation that is consistent with Eq.\ref{eq:true_plant_ode} through numerical integration of the Ordinary Differential Equations (ODEs).

The problem of system identification as described in Section \ref{subsec:SysID} is to find a representative approximation of  Eq.\ref{eq:true_plant}. In this paper, we assume that the $p-$step ahead prediction problem is of vital interest for the purpose of MPC.  A sampling time, $\Delta t$ of $0.1$ (units of time) is used in this paper. 

%%%%%%%%%%%%%%%%%%%%%%%%%%%%%%%%%%%%%%%%%%%%%%%%%%
\subsection{Closed-Loop Control Case Studies} \label{subsec:eval_case}
%%%%%%%%%%%%%%%%%%%%%%%%%%%%%%%%%%%%%%%%%%%%%%%%%%
The approach taken in constructing a RNN for the purpose of MPC is to learn the RNN model with training data, and validate against previously unseen validation/ test data. However, given that the end goal is control, the final evaluation would be in terms of closed-loop performance for the two scenarios, relative to a benchmark/ reference Non-linear MPC (NMPC) approach that uses the true plant model (see ODEs in Eq.\ref{eq:true_plant_ode}) as the internal controller model for on-line control. 

These scenarios are carefully selected and the corresponding conditions are described in Table \ref{tab:cases}. These exemplary scenarios are firstly, a (i) reactor system start-up case and secondly, (ii) an upset-recovery case where the controller needs to recover from an external upset. In the start-up scenario (case I), the system is assumed to be at an initial condition corresponding to a low temperature state that corresponds to relatively low product concentration. In the recovery scenario (case II), the system is assumed to be at an initial condition that corresponds to a region of relatively high temperature state and low yield.

\begin{figure}
 \includegraphics[width=\linewidth]{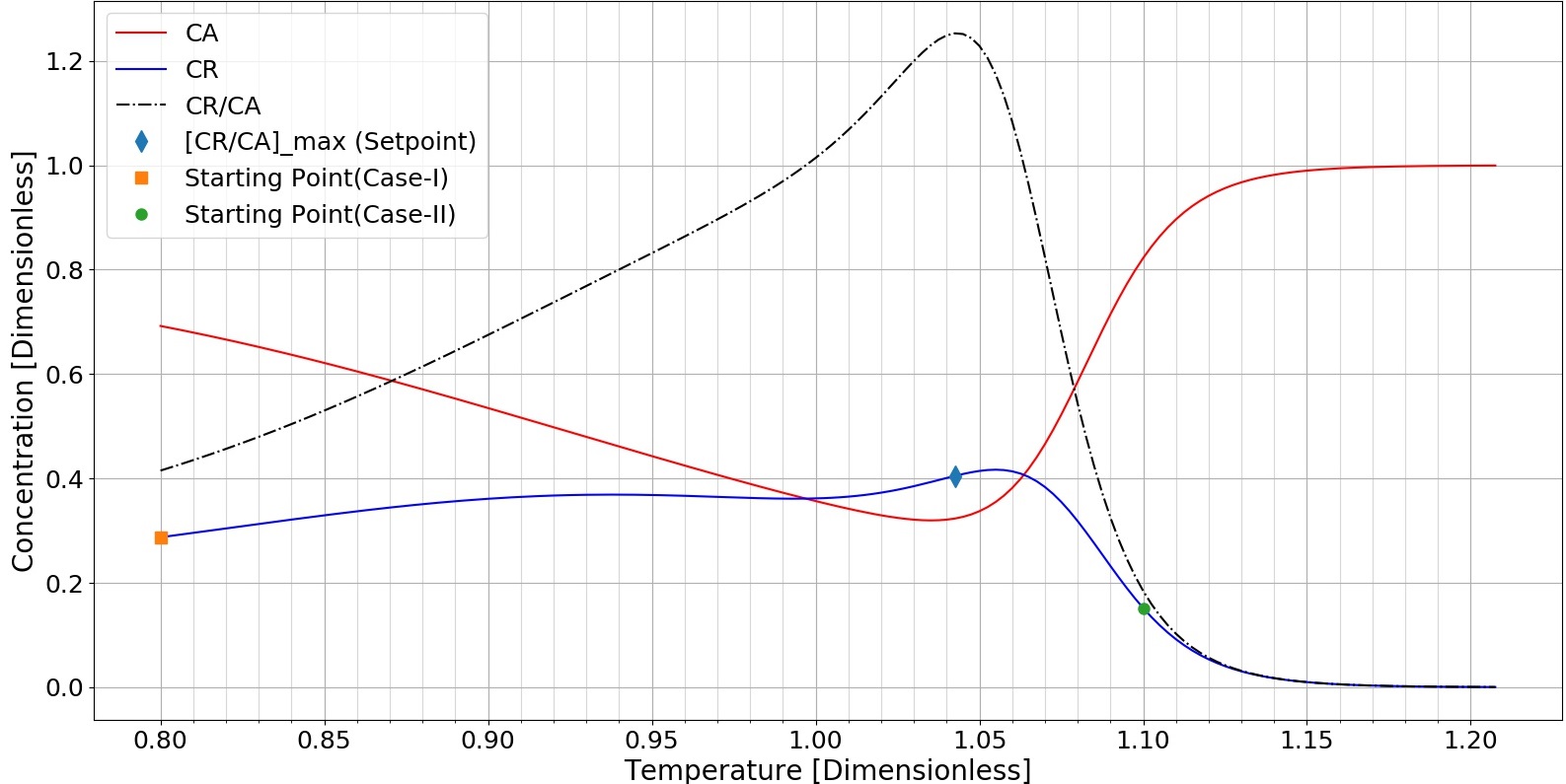} %fig%
 \caption{Starting conditions relative to set-point for case I (start-up) and case II (recovery) \textbf{}}
 \label{fig:initial_conditions}
\end{figure}

The initial condition of the system corresponding to Case I is located on the left of  peak $C_R$, as demarcated by an orange point in Fig. \ref{fig:initial_conditions}. The initial condition of the system corresponding to Case II is located to the right of the peak, as demarcated by the green point in Fig. \ref{fig:initial_conditions}. %These are as shown in Fig. \ref{fig:initial_conditions}. 
% * <xnwang@ucdavis.edu> 2018-05-14T04:43:42.599Z:
% 
% > The initial points corresponding to Cases I and II are to the left and right of  peak $C_R$ 
% This sentence is a bit confusing if readers do not pay enough attention.
% 
% ^ <weechin.wong@gmail.com> 2018-05-17T08:42:27.349Z:
% 
% Expanded the writing to make things clearer.
%
% ^.
%% FIG
The set-point is judiciously chosen to be at a location where the ratio of product concentration to feed concentration is maximum and as is at a product concentration level that is only slightly lower than what is maximally achievable. The rationale is that this operating point maximizes yield whilst lowering downstream separations cost. Table \ref{tab:cases} contains the initial steady-state values of the system as well as the target set-points. 
%% TABLE
\begin{table}[tbp]
\centering
\caption{Steady-state conditions (nominal flow rate, $q$, equals 0.8)}
\label{tab:cases} %tab%
\begin{tabular}{lcccc}
\hline
 \textbf{Scenarios}  	& $C_A$ (Feed) & $C_R$ (Product)  & $q$ (Flow Rate)    	   & $T$ (Temperature)    \\ \hline
(I) Startup   			&   0.692      &  0.287     & 0.800     &   0.800    \\ \hline
(II) Upset recovery  	& 0.822        &   0.152    & 0.800     &  1.100     \\ \hline
Set-point (maximum ratio of $C_R$:$C_A$)    & 0.324 & 0.406 & 0.800 & 1.043 \\ \hline
%Maximum Product Concentration &0.355 & 0.417 & 0.80 & 1.055
\end{tabular} 
\end{table}
%%
%%%%%%%%%%%%%%%%%%%%%%%%%%%%%%%%%%%%%%%%%%%%%%%%%%%%%%%%%%%%%%%%%%%%%%%%%%%%%%%%%%%
\section{Methodologies} \label{sec:SysID}
% * <xnwang@ucdavis.edu> 2018-05-15T17:16:14.402Z:
% 
% > \section{Non-linear Time-Series System Identification via Recurrent Neural Networks} \label{sec:SysID}
% - Please merge Section 3 and 4 to as 3.1-3.2 for a complete "methodology" section
% - The discussion of platform/software used should be both in the next section (currently section 4 talks about the SQP already)
% 
% ^ <weechin.wong@gmail.com> 2018-05-17T08:42:50.471Z:
% 
% Done.
%
% ^ <weechin.wong@gmail.com> 2018-05-17T08:42:57.046Z.
%%%%%%%%%%%%%%%%%%%%%%%%%%%%%%%%%%%%%%%%%%%%%%%%%%%%%%%%%%%%%%%%%%%%%%%%%%%%%%%%%%%

%%%%%%%%%%%%%%%%%%%%%%%%%%%%%%%%%%%%%%%%%%%%%%%%%%%%%%%%%%%%%%%%%%%%%%%%%%%%%%%%%%%
\subsection{Non-linear Time-Series System Identification via Recurrent Neural Networks} \label{subsec:SysID}
% * <xnwang@ucdavis.edu> 2018-05-15T17:16:14.402Z:
% 
% > \section{Non-linear Time-Series System Identification via Recurrent Neural Networks} \label{sec:SysID}
% - Please merge Section 3 and 4 to as 3.1-3.2 for a complete "methodology" section
% - The discussion of platform/software used should be both in the next section (currently section 4 talks about the SQP already)
% 
% ^.
%%%%%%%%%%%%%%%%%%%%%%%%%%%%%%%%%%%%%%%%%%%%%%%%%%%%%%%%%%%%%%%%%%%%%%%%%%%%%%%%%%%
In our case, an MPC based on a single linear model would result in poor control performance due to the presence of input multiplicities corresponding to the change in gain. Thus, there is a need of a fit non-linear model for good closed-loop performance of the associated NMPC controller.

\begin{figure}
\centering
 \includegraphics[width=\linewidth]{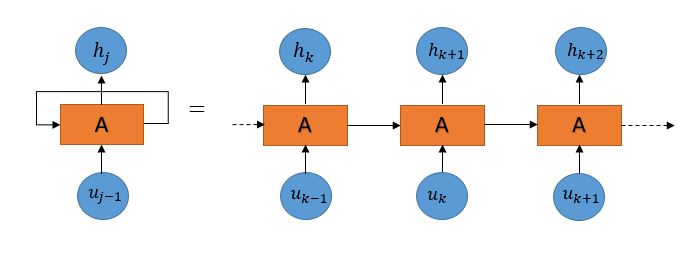}
 \caption{RNN - single cell (left) \& unfolded across time (right)}
 \label{fig:rnn_struct_unrolled} %fig%
\end{figure}

\begin{figure}
\centering
 \includegraphics[width=\linewidth]{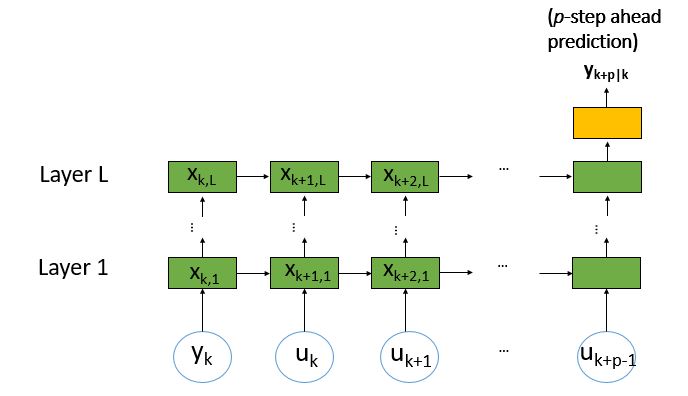}
 \caption{RNN structure for the $p$-step ahead prediction problem}
 \label{fig:rnn_struct} %fig%
\end{figure}

For clarity of exposition, consider a single RNN cell as follows (see Eq.\ref{eq:rnn_state2state}). At time step $k$, the hidden state variable of the RNN is the output of the cell and is denoted by $h_k$; the input to the node is defined as $u_{k-1}$\footnote{While the  machine-learning literature does not account for a time-lag between input state and hidden state, we introduce a time-lag of 1 time-step so that the causal relationship is made explicit.}. 
%%%EQN%%%
\begin{eqnarray}
h_k &=& \sigma\bigg(W_{h,h}h_{k-1}+ W_{u,h}u_{k-1} +b_h  \bigg) \label{eq:rnn_state2state}  
\end{eqnarray} %Eqn.
where $\sigma(.)$ is an activation function applied element-wise, $W_{u,h}$ is the input-to-hidden-state weight matrix; $W_{h,h}$ is the hidden-state-to-hidden-state weight matrix and $b_h$ is a bias (offset) vector. The dimensionality of $h$ and therefore that of $W_{u,h}$, $W_{h,h}$ and $b_h$ depend on the number of nodes of the RNN cell. This single RNN cell can be unfolded across time-steps, as in Fig. \ref{fig:rnn_struct_unrolled} so that the temporal dependencies are clearly illustrated.  Each cell may output directly to one or more measured observation vectors or output into one or more RNN layers above it (see Fig. \ref{fig:rnn_struct}). 
%%%FIG%%%
RNNs may be multi-layered where each subsequent layer captures more nuanced features than the prior layer. 
%It is noted that up to a certain point, increasing layers improves prediction performance subject to the issues of numerical ill-conditioning and / or over-fitting . 

At time step $k$, in order to make a $p$-step ahead prediction of the output variables, $\hat{y}_{k+p|k}$, a RNN, $\mathcal{N(.)}$, is trained for the purpose of modeling the non-linear system dynamics. The corresponding regressors required to estimate this quantity is defined to be $\phi_k := [y_k, u_k,u_{k+1},\ldots,u_{k+p-1}]$. Note that the first element of the regressors is the output measurement at time-step $k$; this is done as a way to provide some form of initialization for the hidden state at the first instance.

The input sequences are introduced in a vertical, layered structure per Fig. \ref{fig:rnn_struct}.
%% FIGURE
% * <xnwang@ucdavis.edu> 2018-05-13T07:40:45.110Z:
% 
% Please revise the original Figure 4. There are red wavy lines below the symbols
% 
% ^ <weechin.wong@gmail.com> 2018-05-14T01:51:40.446Z:
% 
% Done
% 
% ^.
For MPC with a prediction horizon of $p$-steps, it is noted that the following equation is deployed as shorthand to describe the internal RNN-MPC model:
%% EQN
\begin{eqnarray}
\hat{y}_{k+j|k} &=& \mathcal{N}(\phi_{k-p+j}), j \in \{1,2,...,p\} \label{eq:RNN_pred} %eq%
\end{eqnarray}
where the $\hat{(.)}$ symbol makes explicit that an estimate is being computed.

The various weights are estimated by Back-Propagation Through Time (BPTT), which involves an application of the chain rule in the face of minimizing a certain loss function that is related to the prediction error of the model. In our study, the `Mean Absolute Error' is adopted as the loss function. We employ the Adaptive Moment Estimation (ADAM) optimization algorithm \cite{Kingma2014}, a popular stochastic gradient descent-like method with widely applicability, to learn the parameters and weights of the RNN.

The parameters associated with simple RNNs cells are known to be difficult to train due to the gradient-vanishing and gradient-exploding numerical problems. To resolve this, special RNN cell structures such as Long Short-Term Memory (LSTM) cells  have been developed to address this issue.  In the following cases, we used LSTMs to be the cell structure of interest. Details of the LSTM are left to the Appendix for interested readers.

%%%%%%%%%%%%%%%%%%%%%%%%%%%%%%%%%%%%%%%%%
\subsection{Control Problem Formulation} \label{sec:MPCformulation}
%%%%%%%%%%%%%%%%%%%%%%%%%%%%%%%%%%%%%%%%%
The fundamental strategy underpinning MPC is in the selection of a set of future control moves (spanning a control horizon of $m$ steps) that minimizes a certain cost function based on the desired output trajectory over a prediction horizon (spanning $p$ steps). MPC requires a reasonably accurate model, that captures the essential dynamics of the plant under control, so as to predict dynamic behavior multiple-steps ahead. 

The following Receding Horizon Control (RHC) problem (Eq.\ref{eq:MPC_J_func}) is solved via a mathematical optimization program at every time step $k$, whereupon new information in the form of $y_k$ is made available.
%% EQN
\begin{gather} 
\min_{\{\Delta u_k,\Delta u_{k+1},\ldots,\Delta u_{k+m-1}\}} \bigg\{ \sum_{j=1}^p  (\hat{y}_{k+j|k}-y^*)'Q_y(\hat{y}_{k+j|k}-y^*) + \sum_{j=0}^{m-1}  \Delta u_{k+j}' Q_u \Delta u_{k+j} \bigg\}   \label{eq:MPC_J_func} \\
\hat{y}_{k+j|k} = \mathcal{N}(\phi_{k-p+j}), j \in \{1,2,...,p\}  \label{eq:yhat} \\
u_{k+i} \in [u_{\text{min}}, u_{\text{max}}], i \in \{0,1,\ldots,m-1\} \label{eq:ucon} \\
\Delta u_{k+i} \in [\Delta u_{\text{min}}, \Delta u_{\text{max}}], \in \{0,1,\ldots,m-1\} \label{eq:ducon} %eq%
\end{gather}
Here, vector $y^*$ refers to the set-points of $[C_A^*, C_R^*]'$.  $\Delta u_{k+j} \triangleq u_{k+j} -u_{k+j-1}$, refers to the discrete-time rate of change of the manipulated variables, where it is assumed that there is no change in actuator position beyond the control horizon. That is $\Delta u_{k+m} = \Delta u_{k+m+1},\ldots=0$, $\forall k\in \mathbb{Z}^+$. $Q_y \in \mathbb{R}^{n_y\times n_y}$  and  $Q_u \in \mathbb{R}^{n_u\times n_u}$ are square, symmetric, positive semi-definite weighting matrices that serve to penalize deviations from set-point and excessive actuator movement in a quadratic sense, as is typically used in MPC formulations. Constraints are imposed on both the absolute value of the manipulated variable as well as the rate of change, via respective upper and lower bounds, as shown in Eqs.(\ref{eq:ucon})-(\ref{eq:ducon}). 

The aforementioned optimization problem does not, in general possess special structures amenable to global optimality (e.g., convexity). This is a Non-Linear Programming (NLP) problem, which can be solved by modern, off-the-shelf solvers \footnote{For optimization during MPC, an off-the shelf optimizer based on Sequential Quadratic Programming (SQP), a popular method for solving non-linear constrained optimization problems, implemented in Python (\texttt{scipy.optimize.minimize}) was employed.}.
% * <xnwang@ucdavis.edu> 2018-05-15T16:54:24.586Z:
% 
% > (\texttt{scipy.optimize.minimize}) was employed.
% So the optimization option is "Sequential Least SQuares Programming"? Can we move the {scipy.optimize.minimize} to reference instead of text?
% 
% ^ <weechin.wong@gmail.com> 2018-05-17T08:55:21.646Z:
% 
% Moved to footnote. OK?
% 
% ^.
At each time-step the optimal sequence (vector) of future actuator movements  $\{\Delta u^*_{k},\ldots,\Delta u^*_{k+m-1}\}$ are computed whereupon the first element, $\Delta u_k^*$, is implemented and maintained until the next sampling instant, $k+1$. Thereafter, the optimization is repeated in a moving horizon fashion. 

\section{Results and Discussion} \label{sec:resultsandDiscussion}
%%%%%%%%%%%%%%%%%%%%%%%%%%%%%%%%
The procedure for investigating the RNN-MPC approach as a practical control strategy is according to the following steps:
(1) Gathering process operating data with perturbation in all the manipulated variables to generate a rich-enough data set. %Certain data treatment method such as removal of outliers and scaling is recommended.The data should be properly separated into training sets against test sets. 
(2) Training RNN models with various parameters using training sets and validating them with test data. Select the one with best performance that can capture most of the behavior of the plant.
(3) Implementing the chosen RNN model into a properly designed MPC for the final purpose of control. %Using some stress tests to check the robustness of the RNN-MPC model to decide whether a better RNN model is necessarily needed. 

%%%%%%%%%%%%%%%%%%%%%%%%%%%%
\subsection{System Identification Results} \label{subsec:SysIDResults}
\begin{figure}
 \includegraphics[width=\linewidth]{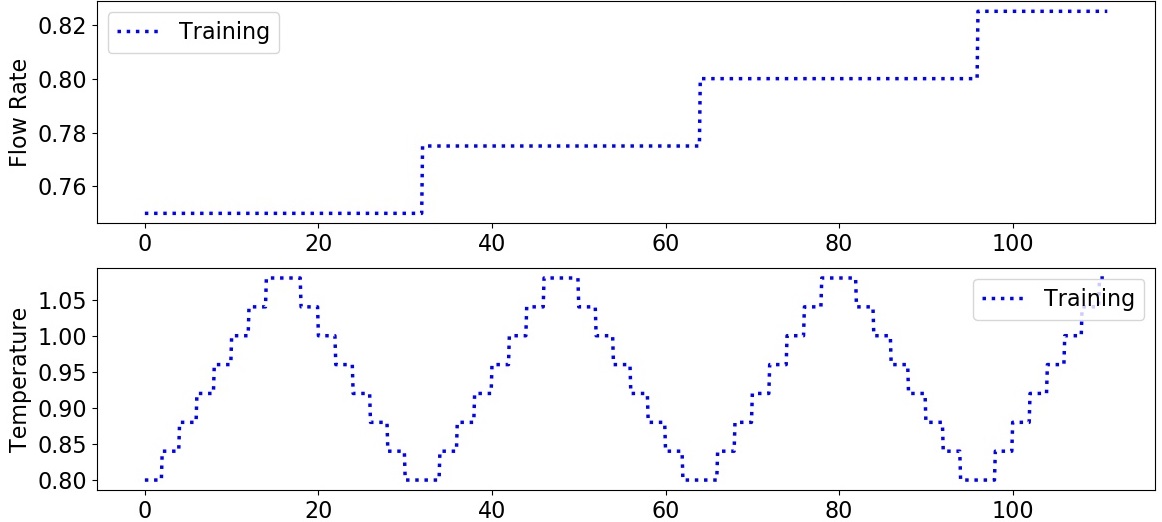} %fig%
 \caption{Training data for system identification (Manipulated Variable; sampling rate, $\Delta_t = 0.1$)}
 \label{fig:sysID_trg_data_mv}
\end{figure}

\begin{figure}
 \includegraphics[width=\linewidth]{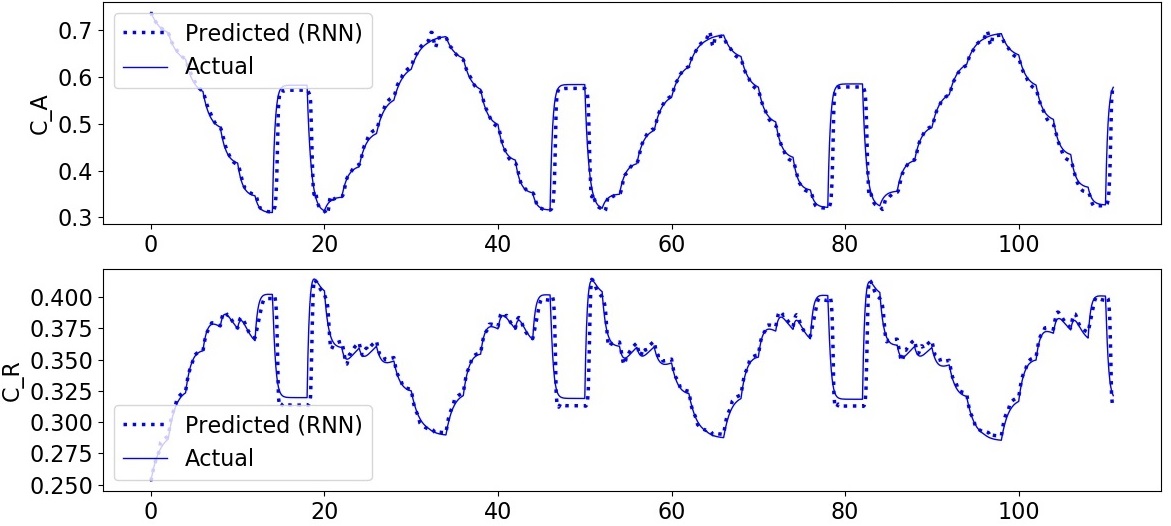} %fig%
 \caption{System identification - model $p$-step ahead prediction on training data [1000 nodes, $2$ layers]}
 \label{fig:sysID_fit_trg_output}
\end{figure}

%%%%%%%%%%%%%%%%%%%%%%%%%%%%
Using Eq.\ref{eq:true_plant_ode} to simulate the true plant behavior, a profile of the manipulated variables per Fig. \ref{fig:sysID_trg_data_mv} was used to generate the necessary training data for the RNN model. Reactor temperature was stepped up (past the point of inflexion) and then back down for various choices of flow rates as can be seen in Fig. \ref{fig:sysID_trg_data_mv}. The corresponding output profile for RNN can be found in Fig. \ref{fig:sysID_fit_trg_output}.
%%FIG
We varied the number of nodes between $250$ to $2,000$ and number of layers to be between $1$ and $3$. Table \ref{tab:rnn_test_data_summary} summarizes the performance of the various RNN parameters over previously unseen test (validation) data. System identification performance over $N$ data points is quantified by Root Mean Square Error (RMSE). 
\begin{eqnarray}
RMSE&=& \sqrt{ \left(\frac{1}{N-p+1}\sum_{k=0}^{N-p}\{y_{k+p}-y_{k+p|k}\}^2 \right) }
\end{eqnarray}
%
%% TABLE
\begin{table}[!tbph]
\centering
\caption{RMSE of RNN model over test data ($1,000$ training epochs)}
\label{tab:rnn_test_data_summary} %tab%
\begin{tabular}{cccc}
\hline\
 No. Layers / No. Nodes	& \textbf{250} & \textbf{500} & \textbf{1000}   \\ \hline
	\textbf{1}   	& 0.0299 		&  0.0268 &  0.0206  		 \\ \hline
	\textbf{2}     & 0.0238  		&  0.0118 &   0.0083		 \\ \hline
	\textbf{3}  	& 0.0262  		&   0.0119 & 0.0125  		 \\ \hline
\end{tabular} 
\end{table}
Generally speaking, the prediction performance is found to improve with increasing number of nodes. This is because more expressive features are created when increasing number of nodes to better fit the nonlinear dynamic system behavior, thereby reducing prediction bias. 

Furthermore, a deep, hierarchical model may be much more efficient in representing some functions than a shallow one \cite{Pascanu2013}. We observed  this  when the number of layers were increased from $1$ to $2$. However, performance deteriorated when the number of layers were increased from $2$ to $3$.  

This is  due to the fact that RNNs that are too deeply stacked may be difficult to learn, without extensive tuning potentially with heuristics and/ or specially developed algorithms. This is potentially due to the aforementioned gradient vanishing numerical problems that arise during back-propagation. This issue is attributable to the complex structure of multi-layered RNN. %(Ziv Aharoni, Gal Rattner, Haim Permuter).

As can be seen in Table \ref{tab:rnn_test_data_summary}, using $1,000$ nodes per RNN (LSTM) cell with $2$ hidden RNN layers was found to yield the best prediction performance over previously unseen test data. Consequently, we opted to proceed in our system identification studies, and corresponding closed-loop control studies, by constraining the number of RNN layers to $2$. For completeness, we also proceeded to learn an RNN model with $2$ layers and $2,000$ nodes. We noticed that over-fitting occurred, as reflected in Table\ref{tab:rnn_test_data_summary2}.

\begin{figure}
 \includegraphics[width=\linewidth]{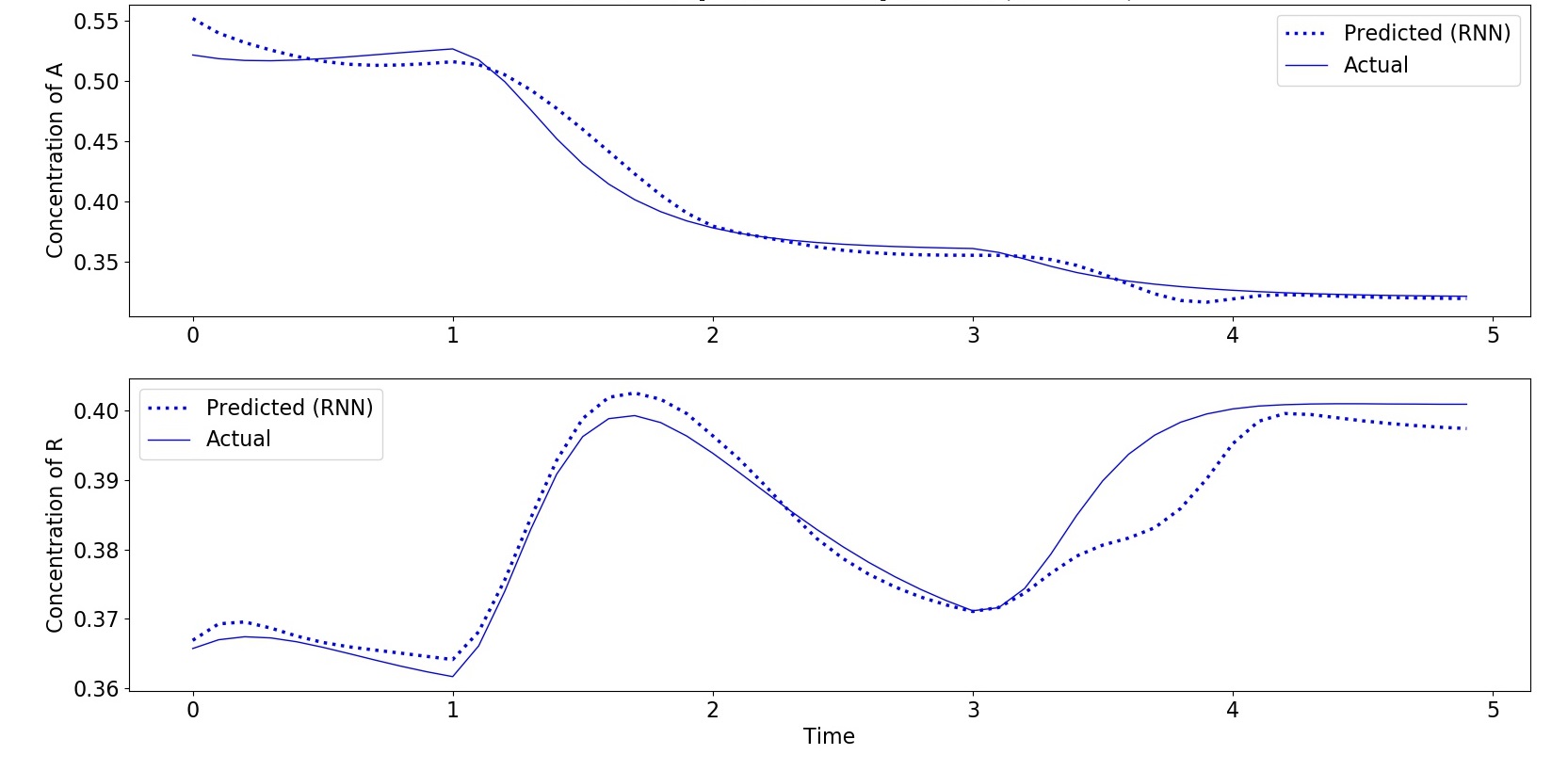} %fig%
 \caption{Model $p$-step ahead prediction on test data for system identification [1000 Nodes, 2 Layer]}
 \label{fig:sysID_fit_test_output}
\end{figure}

% * <xnwang@ucdavis.edu> 2018-05-15T17:05:05.229Z:
% 
% > For completeness, we learned a 2-layer, 250-node RNN model using 250 nodes, with the hope of improving the performance over the 1-layer, 250-node situation that suffers from under-fitting. In this event, we see a deterioration of prediction results, which is contrary to the theoretical expectation. This supports our decision to proceed with single-layer RNN solutions.
% This paragraph sounds like a "reporting" tone. Why does it obey the theoretical expectation? We don't have to say this if no explanation of the performance variations is provided. 
% 
% ^ <weechin.wong@gmail.com> 2018-05-17T09:01:39.881Z:
% 
% We changed it to "general" expectation, as stated in line 351, as opposed to "theoretical expectation".  Also, expanded the system identification part to study the effect of more layers. Hope this makes clear and also enriches the study.
% 
% ^.
%% TABLE
\begin{table}[!tbph]
\centering
\caption{Performance of RNN model over test data for $2,000$ nodes and $2$ layers ($1,000$ training epochs). Results for $1,000$ nodes and 2 layers are reproduced for ease of reference.}
\label{tab:rnn_test_data_summary2} %tab%
\begin{tabular}{ccc}
\hline\
 No. Layers	& No. Nodes & RMSE (Test-Data)   \\ \hline
   2       &1000 & 0.0083 \\ \hline
    2       &2000 & 0.0177 \\ \hline
\end{tabular} 
\end{table}
%%
\begin{comment}
%% TABLE
\begin{table}[!tbph]
\centering
\caption{RMSE of RNN model over test data ($1,000$ training epochs)}
\label{tab:rnn_test_data_summary} %tab%
\begin{tabular}{cccc}
\hline\
 No. Layers / No. Nodes	& \textbf{250} & \textbf{500} & \textbf{1000}   \\ \hline
	\textbf{1}   	& 0.0299 		&  0.0268 &  0.0206  		 \\ \hline
	\textbf{2}     & 0.0238  		&  0.0118 &   0.0083		 \\ \hline
	\textbf{3}  	& 0.0262  		&   0.0118 & 0.0125  		 \\ \hline
\end{tabular} 
\end{table}
%%
\end{comment}
The corresponding prediction performance for the best RNN on the training data ($1,000$ nodes and $2$ layers) is shown in Fig. \ref{fig:sysID_fit_trg_output} and that on the test data can be seen in Fig. \ref{fig:sysID_fit_test_output}. For the case of the training data, it can be seen that there is very good fit. With highly dynamic input data, the resulting highly dynamic output information can be captured well. For the previous unseen test data, this well tuned RNN can show a good fit which captures all the general trend of the output information by given highly dynamic input data. Although the performance of the RNN for system identification of the test data is not perfect, it shows very promising result when implemented in MPC for closed-loop control purpose. The closed-loop results for $2$ layers are shown in the following discussions\footnote{Code implementation is performed using the Python platform (ver 3.6.5) with Tensorflow (ver 1.7.0), equipped with a Keras (ver 2.1.5) interface for RNN learning. Numerical integration, where necessary, was performed using \texttt{scipy.integrate.ode} with a sample rate of 0.1 (time units).}.
%% Fig: Sys Id Training Data
%
%% Fig: Sys Id Test Data
%
%%%%%%%%%%%%%%%%%%%%%%%%%%%%%%%%%%%%%%%%%%%%%%%%%%%%%%%%%%%%%%
\subsection{Closed Loop RNN-MPC Performance}  \label{subsec:CL_Results}
%%%%%%%%%%%%%%%%%%%%%%%%%%%%%%%%%%%%%%%%%%%%%%%%%%%%%%%%%%%%%%
With reference to Fig. \ref{fig:initial_conditions}, for both cases, the objective is to have the system converge to a target set-point that corresponds to an operating point where the ratio of $R$ concentration against that of $A$ is at its maximum, thereby optimizing yield and potential downstream separation costs. It is noted that Linear MPC, based on a single linear model, will not be able to perform well \cite{Bequette2007}, thereby necessitating the selection of a non-linear MPC solution. 

For both scenarios, the following MPC controller parameters are used: $p=m=1.0$   (time units),  $Q_y=\text{diag}[2.4,5.67]$, $Q_u=\text{diag}[25,25]$. Total simulation time times,$N$, is set to $40$.

The elements of $Q_y$ are selected to reflect that tracking $C_R$ is more important than tracking $C_A$. The $Q_u$ penalty on excessive actuator movement $\Delta u$ is higher than that on the state variables so as to avoid over-aggressive controller actions, which may lead to system instability. 

For digital implementation, the sampling rate is $\Delta t=0.1$ [\text{dimensionless time units}].  We assumed the flow rate and temperature to be constrained such that $q\in [0.75,0.85]$, $T \in [0.5,1.1]$. Also, the rate of change of the manipulated variables were constrained to be as such: $\Delta q \in [-0.1,0.1]$ and $\Delta T \in [-0.1,0.1]$, where the $\Delta$ operator represents time-differencing.

In the following studies, we perform closed-loop controller performance benchmarking. This is done against a NMPC controller where the controller's internal model is the same as the actual, full plant model (as in Eq. \ref{eq:true_plant}). The optimizer / solver deployed for both the RNN-MPC and benchmark NMPC methods remain the same as the one mentioned in the Section \ref{sec:MPCformulation}.

In comparing controller performance, the index used is naturally, the sum of the stage-wise MPC cost added over the total number of discrete-time steps, $N$, across the entire simulation horizon:
\begin{eqnarray}
J&=&  \sum_{k=1}^N \bigg\{ (y_{k}-y^*)'Q_y({y}_{k}-y^*) +  \Delta u_{k}' Q_u \Delta u_{k} \bigg\}  \label{eq:J_idx} \\
\mathcal{I} &=& \left\{ 1- \tfrac{J_{[\text{RNN}]}-J^*}{J^*} \right\}*100 \%
\end{eqnarray}
where $Q_y$, $Q_u$ have already been defined in Eq.\ref{eq:MPC_J_func} 
for the benchmark NMPC. The sequence of controller actions in Eq.\ref{eq:J_idx} are determined per the corresponding controllers. Specifically, 
$J_{[\text{RNN}]}$ refers to the corresponding total stage-wise cost for the RNN-MPC solution and $J^*$ refers to that for the benchmark NMPC.

For both scenarios, the system is allowed to run for $1.0$ time units (equivalent to the prediction horizon) before the controller is switched on. This is because the RNN-MPC is required to store in its memory a history of $p$ number of manipulated input variables.

Figures \ref{fig:cl_N250}-\ref{fig:cl_N1000} below and \ref{fig:cl_N500}-\ref{fig:cl_N2000} in the Appendix show the closed-loop time-series for both scenarios as a function of different numbers of nodes. As can be seen from these plots, the RNN-MPC approach performs well for both complex control scenarios, in general. For example, all controllers exhibited stability, even in the under-fitted case of 250 nodes. This suggests a certain robustness associated with this combination. Table \ref{tab:cl_rnn_mpc} summarizes a quantification of RNN-MPC performance by way of the closed loop performance index, averaged over both scenarios, for ease of comparison.

In the case of $250$ nodes (see Fig. \ref{fig:cl_N250}), it can be seen that a significant offset occurs at steady-state for both scenarios even though initial transient performance is somewhat similar to that of the benchmark NMPC solution.
%In the case of $500$ nodes (see Fig.  \ref{fig:cl_N500}), closed loop performance has improved over the $250$ nodes case. However, there does exist some observable offset in scenario 2 (i.e., Fig.  \ref{fig:N500_L1_recovery}).
$1,000$ nodes yielded the best closed-loop performance as evident in Fig. \ref{fig:cl_N1000}. This is consistent with the previous observations made during system identification. Steady-state tracking with no offset occurs for both scenarios. While the closed-loop trajectory of the RNN-MPC solution is close to that of the benchmark solution and has same total stage-wise cost as the benchmark NMPC solution. It is noted that in scenario $1$, the transient performance is slightly more rapid than that of the benchmark. This is because the actuator movement, as informed by the SQP optimizer, is more aggressive (though still within the imposed constraints). In the case of $2,000$ nodes (see Appendix Fig. \ref{fig:cl_N2000}), closed loop performance worsens as a result of over-fitting. We summarize the aforementioned discussion in Table \ref{tab:cl_rnn_mpc}, from which it is clear that $1,000$ nodes yielded the best closed-loop performance.

%% SUMMARY TABLE of CL-CTRL
\begin{table}[!htbp]
\centering
\caption{Performance of closed-loop RNN-MPC as a function of RNN nodes (2 layers)}
\label{tab:cl_rnn_mpc} %tab%
\begin{tabular}{ccc}
\hline\
 No. Nodes & average performance index, $\mathcal{I}_{\text{avg}}$ & Comments \\ \hline
	250  		&   93.7    		&Steady-state Offset \\ \hline
	500  		&   95.8    		&Steady-state Offset\\ \hline
	1000 		&   100.0    		& Desired Performance\\ \hline
    2000        &   98.6  			&Steady-state Offset \\
\end{tabular} 
\end{table}
%%
\begin{comment}
%% SUMMARY TABLE of CL-CTRL
\begin{table}[!htbp]
\centering
\caption{Performance of Closed-Loop RNN-MPC as a function of RNN Nodes}
\label{tab:cl_rnn_mpc} %tab%
\begin{tabular}{ccc}
\hline\
 No. Nodes & Performance Index - Average  & Comments \\ \hline
	250  		&   93.7 (93.2,94.1)    		&Steady-state Offset \\ \hline
	%2  		& 250  		&   90.7 (81.5,100.0)       &Offset\\ \hline
	500  		&   95.8 (100, 91.6)   		&Steady-state Offset\\ \hline
	%2  		& 500  		&   97.2 (94.3,100.0)   	&Offset \\ \hline
	1000 		&   100.0 (100.0,100.0)   	&- \\ \hline
	%2  		& 1000 		&   99.5 (100.0,99.0)   	& Offset \\ \hline
    2000        & 98.6 (97.1,100.0) &Steady-state Offset \\
\end{tabular} 
\end{table}
%%
\end{comment}
%%%%%%%%%%%%%%%%%%%%
%%250 NODES; 1 LAYER
%%%%%%%%%%%%%%%%%%%%
%
\begin{figure}[!htbp]
\begin{subfigure}{1.00\textwidth}
\includegraphics[width=\linewidth]{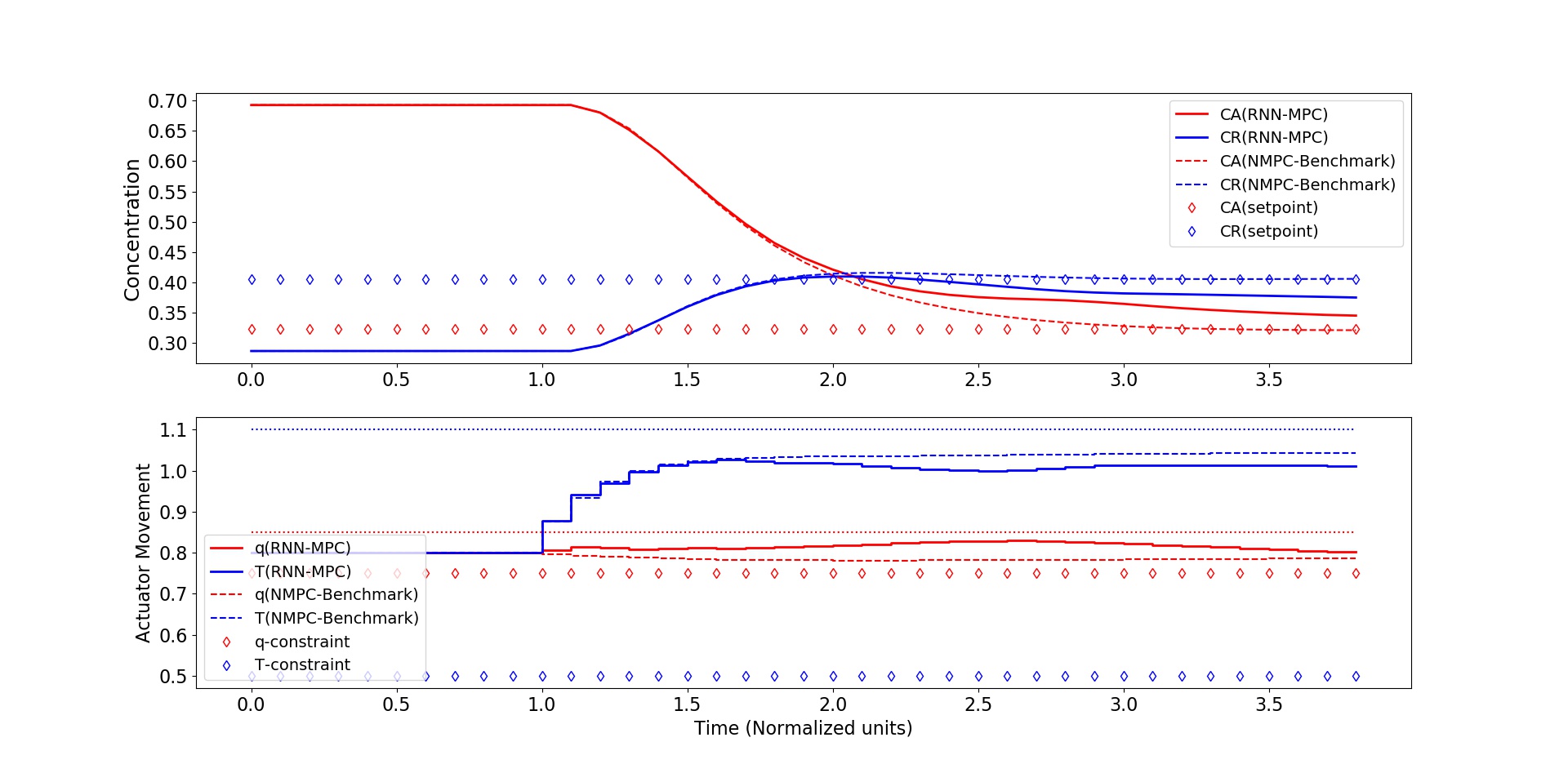} 
\caption{250 nodes; 2 RNN layers - startup}
\label{fig:N250_startup}
\end{subfigure}
\begin{subfigure}{1.00\textwidth}
\includegraphics[width=\linewidth]{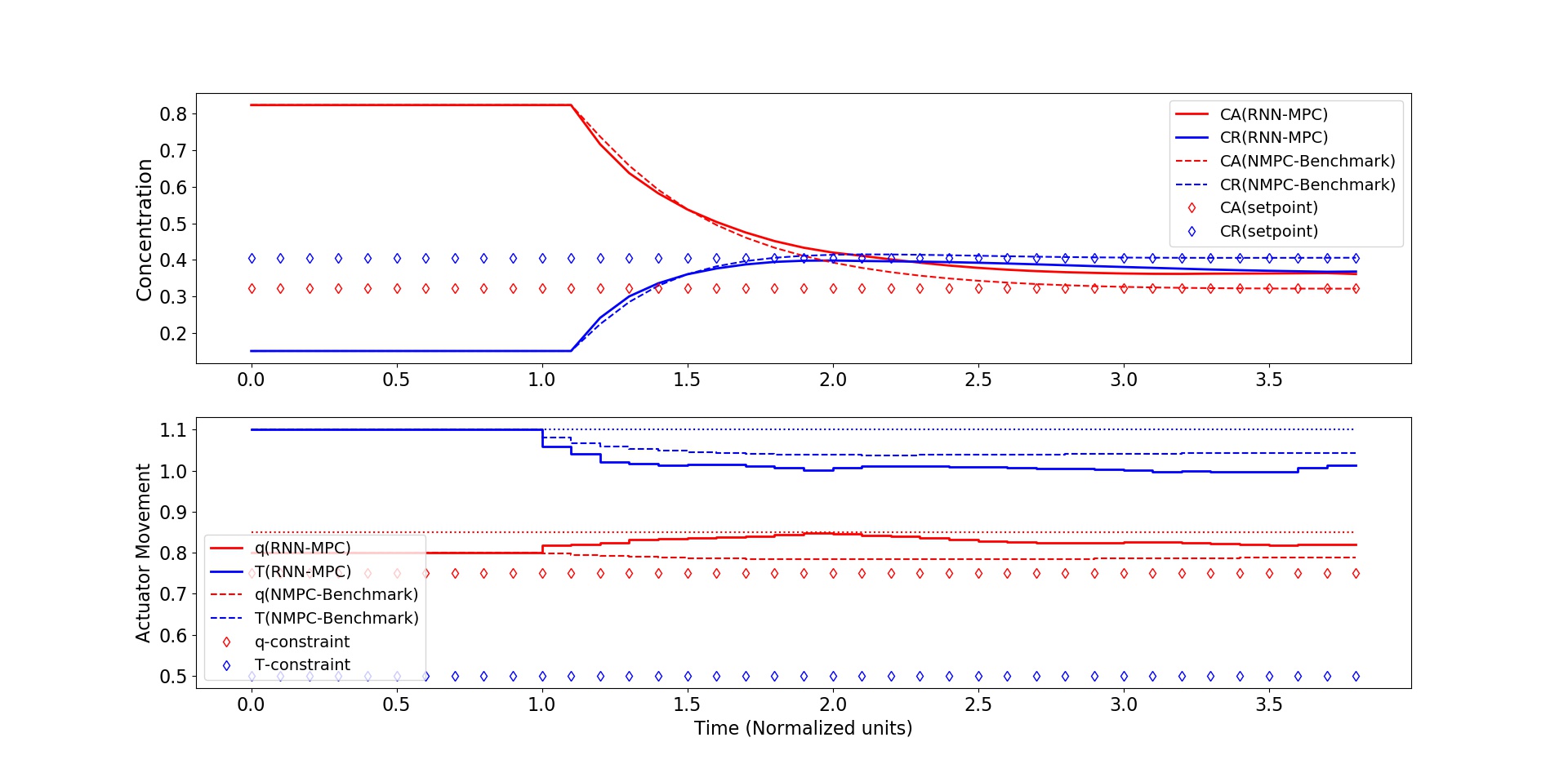}
\caption{250 nodes; 2 RNN layers - recovery}
\label{fig:N250_recovery}
\end{subfigure}
\caption{Closed-loop performance with 250 nodes, 2 layers}
\label{fig:cl_N250}
\end{figure}
%
%%%%%%%%%%%%%%%%%%%%
%%1000 NODES; 1 LAYER
%%%%%%%%%%%%%%%%%%%%
%
\begin{figure}[!htbp]
\begin{subfigure}{1.00\textwidth}
\includegraphics[width=\linewidth]{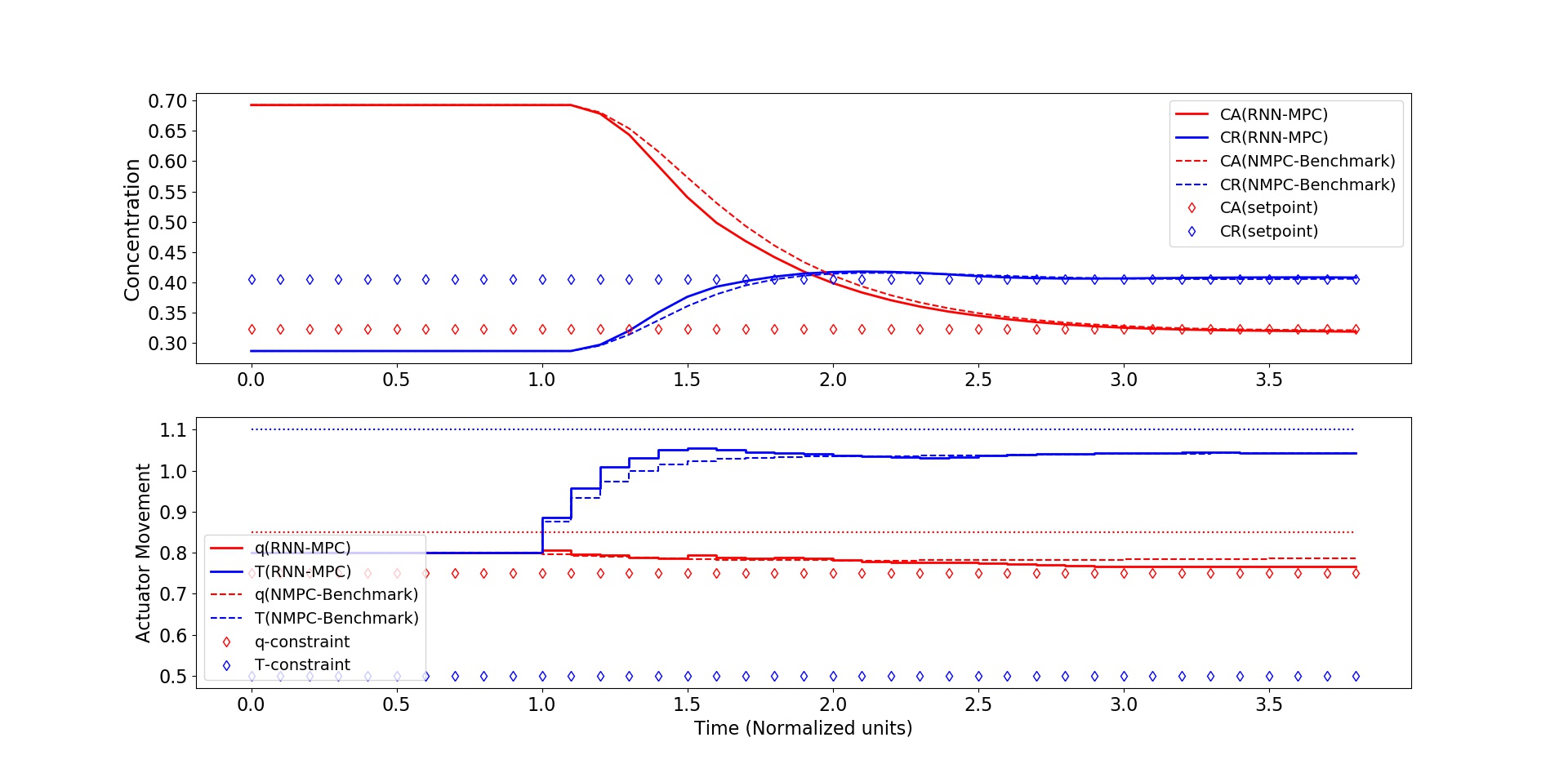} 
\caption{1000 Nodes; 2 RNN layers - startup}
\label{fig:N1000_startup}
\end{subfigure}
\begin{subfigure}{1.00\textwidth}
\includegraphics[width=\linewidth]{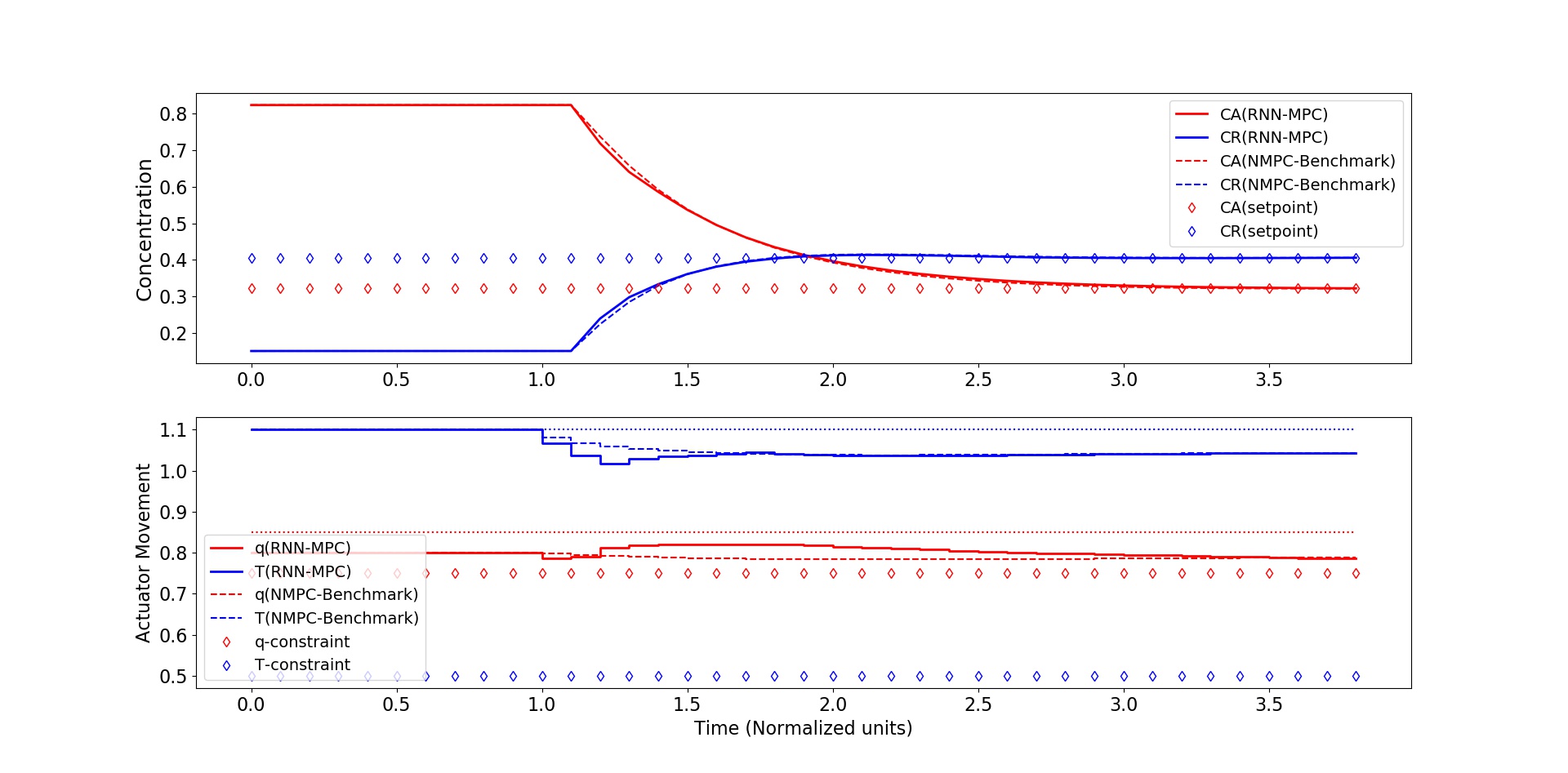}
\caption{1000 Nodes; 2 RNN layers - recovery}
\label{fig:N1000_recovery}
\end{subfigure}
\caption{Closed-loop performance with 1000 nodes, 2 layers}
\label{fig:cl_N1000}
\end{figure}
%

\begin{comment}
A table of the RMSE of open-loop system identification results against closed-loop performance index was constructed. 
%%
\begin{table}[!htbp]
\centering
\caption{Trend -  Test Data Prediction Error (RMSE) vs. Closed-Loop Performance (Averaged Over Scenarios)}
\label{tab:rmse_vs_cl} %tab%
\begin{tabular}{cc}
\hline\
    RMSE					& Avg. Closed-Loop Performance Index    \\ \hline
	0.0083  				& 100.0		   \\ \hline
	0.0119					& 95.8  	   \\ \hline
	\textcolor{red}{0.0177}	& 98.6  	   \\ \hline
	0.0238					& 93.7 		   \\ \hline
\end{tabular} 
\end{table}
\end{comment}
%Apart from this, the closed-loop performance per fig and table shows that even with fairly poor performance RNN model (500 nodes, 1 layer) the RNN-MPC approach can tackle with both complex control scenarios with only small steady-state offset. 

%In addition, the good-fit RNN model (1000 nodes, 1 layer) has excellent close-loop control performance without any offset and has same total stage-wise cost as the bench-marking NMPC. 

In summary, these results illustrate that even if the open-loop system identification does not perfectly describe the actual true plant phenomenon, as long as the general trend and properties are captured, a RNN-MPC approach can lead to a good close-loop control performance. Under a practical situation, a perfect model is infeasible to get normally, our RNN-MPC approach can tackle with this situation well to get good control performance which shows a promising way to deal with real-world continuous pharmaceutical manufacturing challenges. 

\section{Conclusion \& Future Research} \label{sec:conclusion}
%%%%%%%%%%%%%%%%%%%%%%%%%%%%%%%%%%%%%%%%%%%%%%%%%%%%%%%%%%%%%%
We considered an exemplary problem for continuous pharmaceutical manufacturing by studying a single, multi-input multi-output CSTR example (per \cite{Seki2004}, \cite{Koppel1982}) which experiences input multiplicity due to reversible kinetics ($A \leftrightarrow R \leftrightarrow S$). This set-point is close to a point of inflexion where the system gain changes in sign with respect to reactor temperature. 

We show how RNNs can be learned and present associated closed-loop performance results for two scenarios that require the RNN-based NMPC to move from either side of the inflexion point towards the desired set-point. As a result of input multiplicity, it is noted that a single linear controller will not work well for the problem of concern. While favorable results of the RNN-MPC controller are obtained when compared against a NMPC benchmark that uses the true plant model for control in terms of closed-loop performance.

Future research involves extending the methodology to multiple CSTRs and to reaction kinetics of increasing complexity and immediate relevance to the pharmaceutical industry.

\section*{acknowledgements}
We would like to acknowledge Associate Professor Saif A. Khan in the Department of Chemical and Biomolecular Engineering, National University of Singapore for his input related to reaction kinetics and multiple discussions. Also, our thanks to the Pharma Innovation Programme Singapore (PIPS), the source of motivation for this work.
%Acknowledgements should include contributions from anyone who does not meet the criteria for authorship (for example, to recognize contributions from people who provided technical help, collation of data, writing assistance, acquisition of funding, or a department chairperson who provided general support), as well as any funding or other support information.

\section*{conflict of interest}
We declare no conflict of interest.

% Submissions are not required to reflect the precise reference formatting of the journal (use of italics, bold etc.), however it is important that all key elements of each reference are included.

%\bibliography{rnn_paper,rnn_paper2,Mendeley_Pharma_AICHE_Abstract}
%\bibliography{rnn_paper,rnn_paper2,rnn_paper3}
%\bibliographystyle{Author-year}

\newpage
%%%%%%%%%%%%%%%%%%%%%%%%%%%%%%%%
\section*{Appendix} \label{sec:Appendix}
%%%%%%%%%%%%%%%%%%%%%%%%%%%%%%%%
\subsection*{Kinetic Parameters for Plant Model}
$C_{A0}$ refers to the feed concentration of $A$ and is assigned a value of 0.8. The vector of Arrhenius pre-exponentials $k_{0}$ are sequentially assigned  values of $[1.0, 0.7, 0.1, 0.006]$. \\
The normalized activation energies $-\frac{E}{RT_0}$ have values of $[8.33, 10.0, 50.0, 83.3]$.

\subsection*{Long Short-Term Memory Cells} %http://colah.github.io/posts/2015-08-Understanding-LSTMs/
In the conventional RNN structure, back-propagation through time results in the gradient signal being multiplied numerous times viz-a-viz the weights corresponding to the various connections (edges). Depending on the eigen-value of the associated weighting matrix, this leads to the either vanishing (eigen-value less than unity) or exploding (eigen-value greater than unity) gradients. This has the potential to severely impact the learning quality.

The machine-learning community has been using LSTM cells to replace the conventional RNN cell in order to mitigate this issue. These LSTM memory cell uses several gating functions that, in conjunction, serve to adjust / modulate the propagation of signals between cells (e.g., through `forgetting', `input', `output' gates) to avoid said numerical issue.

The basic LSTM cell structure is as follows:
\begin{eqnarray}
h_k &=& o_k*\tanh{\big(C_k\big)} \label{eq:output_value}\\
C_k&=& f_k*C_{k-1} + i_k*\tilde{C}_k \label{eq:cell_state} \\
\tilde{C}_k &=& \tanh{\big( W_C[h_{k-1},u_{k-1}]'+b_c\big)} \label{eq:candidate_value}\\
i_k &=& \sigma \big(W_i.[h_{k-1},u_{k-1}]' + b_i \big) \label{eq:input_gate} \\
f_t &=& \sigma(W_f.[h_{k-1},u_{k-1}]'+ b_f)  \label{eq:forget_gate} \\
o_t&=& \sigma \big(W_o.[h_{k-1},u_{k-1}]' + b_o\big) \label{eq:output_gate}
\end{eqnarray}
where $k$ is the time index, $h_k$ the hidden state variable, $u_k$ the manipulated / input variable. $f_k \in [0,\textbf{1}]$, $i_k \in [0,\textbf{1}]$ and $o_k \in [0,\textbf{1}]$ are termed the `forgetting', `input' and `output' gates respectively. The input gate (Eq.\ref{eq:input_gate}) controls the degree to which the state of the memory cell is affected by the candidate information (Eq.\ref{eq:candidate_value}); the output gate (Eq.\ref{eq:output_gate}) controls how the state of the cell affects other neurons; the forget gate (Eq.\ref{eq:forget_gate}) modulates the cell's self-recurrent connection, allowing the cell to (partially) remember the previous state, similar to traditional RNNs.

These work to combine old ($C_{t-1}$) and new candidate information ($\tilde{C}_t$), respectively, to be passed to subsequent LSTM cells. $\sigma$ is an  activation function that returns a value between $0$ and $\textbf{1}$. $*$ refers to a point-wise multiplication.

\newpage
\subsection*{Additional Figures - Closed Loop Performance}

%%%%%%%%%%%%%%%%%%%%
%%500 NODES; 1 LAYER
%%%%%%%%%%%%%%%%%%%%
%
\begin{figure}[!htbp]
\begin{subfigure}{1.00\textwidth}
\includegraphics[width=\linewidth]{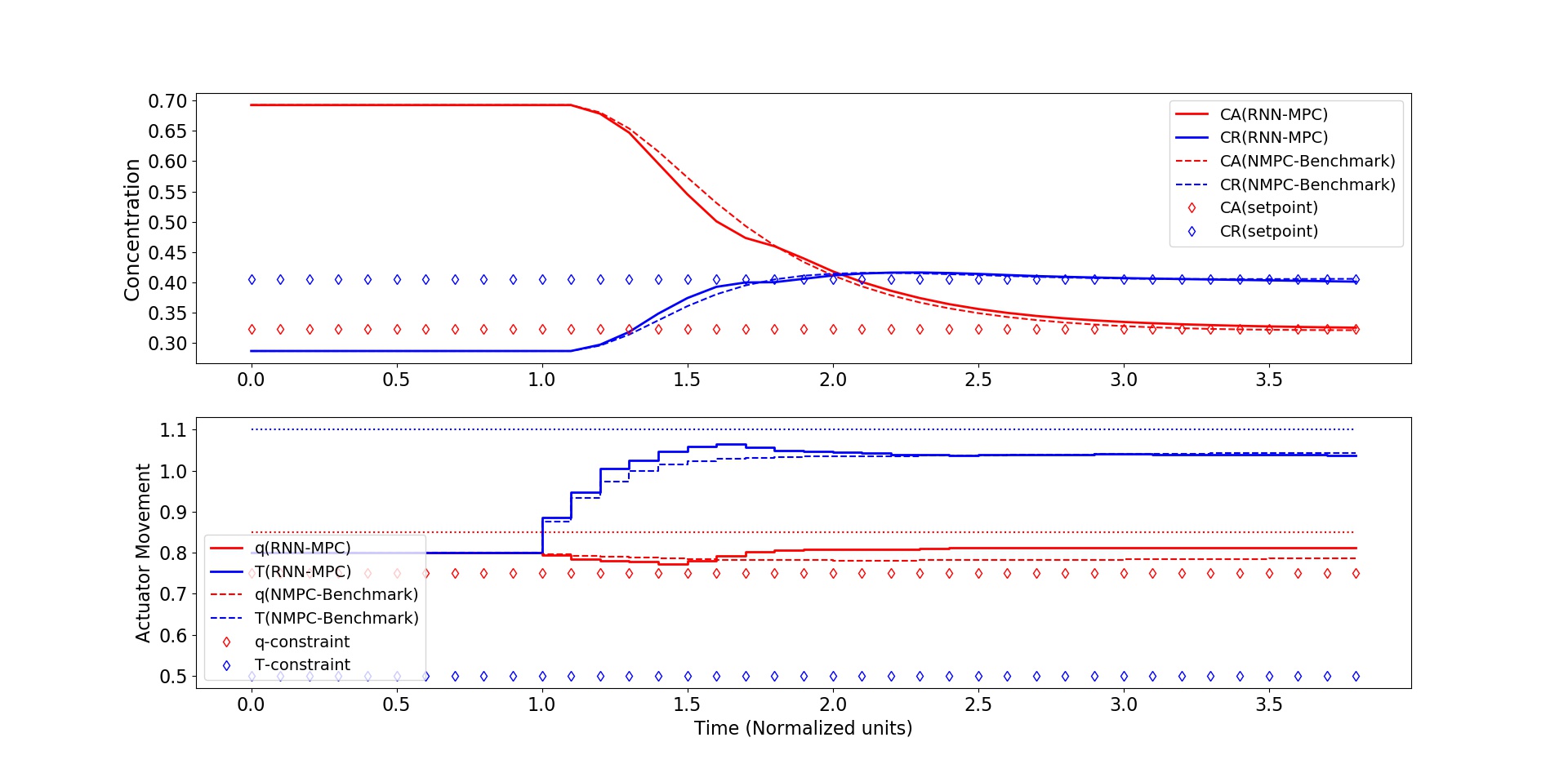} 
\caption{500 nodes; 2 RNN layer - startup}
\label{fig:N500_startup}
\end{subfigure}
\begin{subfigure}{1.00\textwidth}
\includegraphics[width=\linewidth]{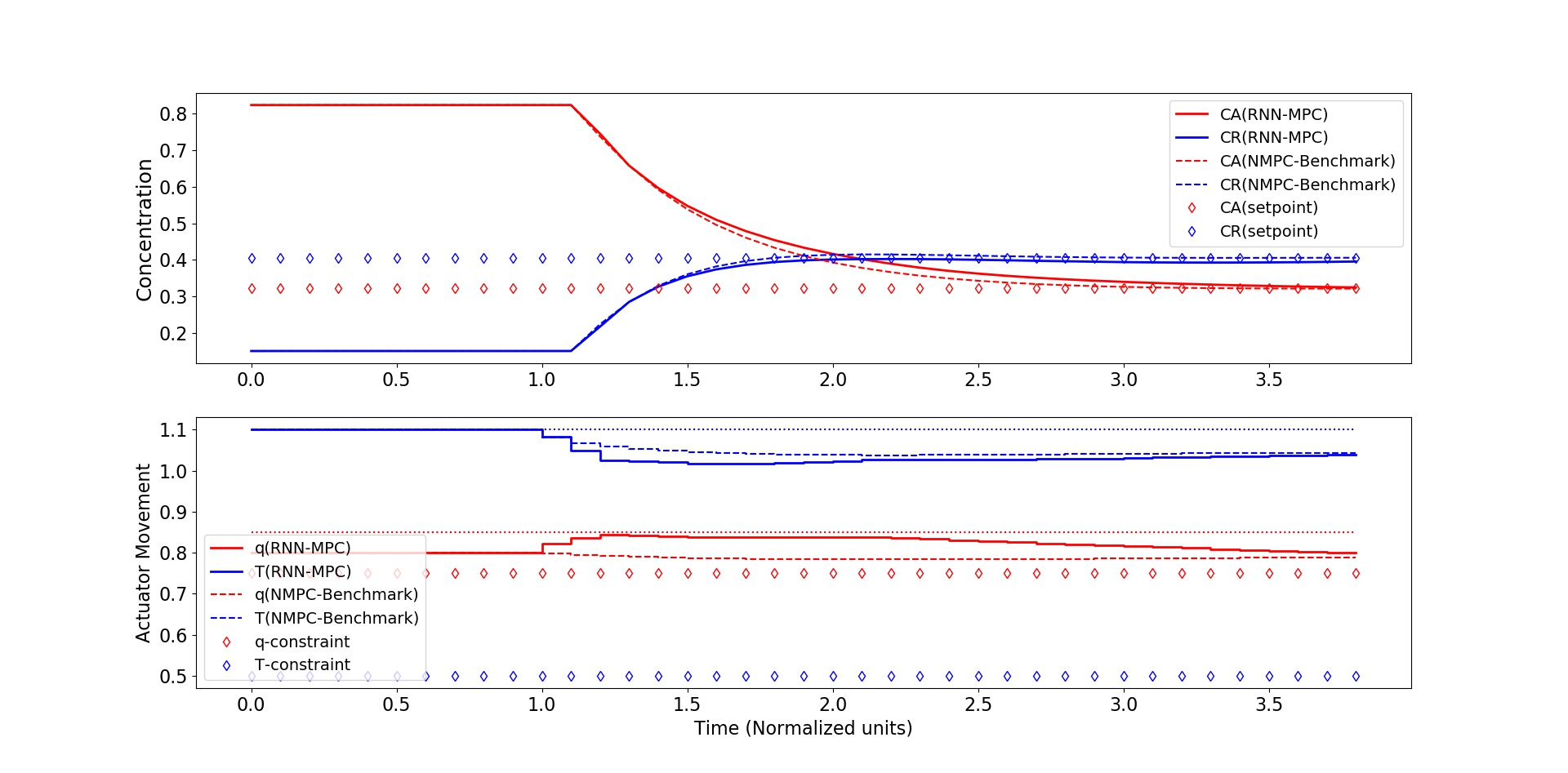}
\caption{500 nodes; 2 RNN layer - recovery}
\label{fig:N500_recovery}
\end{subfigure}
\caption{Closed-loop performance with 500 nodes, 2 layers}
\label{fig:cl_N500}
\end{figure}
%
%%%%%%%%%%%%%%%%%%%%
%%2000 NODES; 1 LAYER
%%%%%%%%%%%%%%%%%%%%
%
\begin{figure}[!htbp]
\begin{subfigure}{1.00\textwidth}
\includegraphics[width=\linewidth]{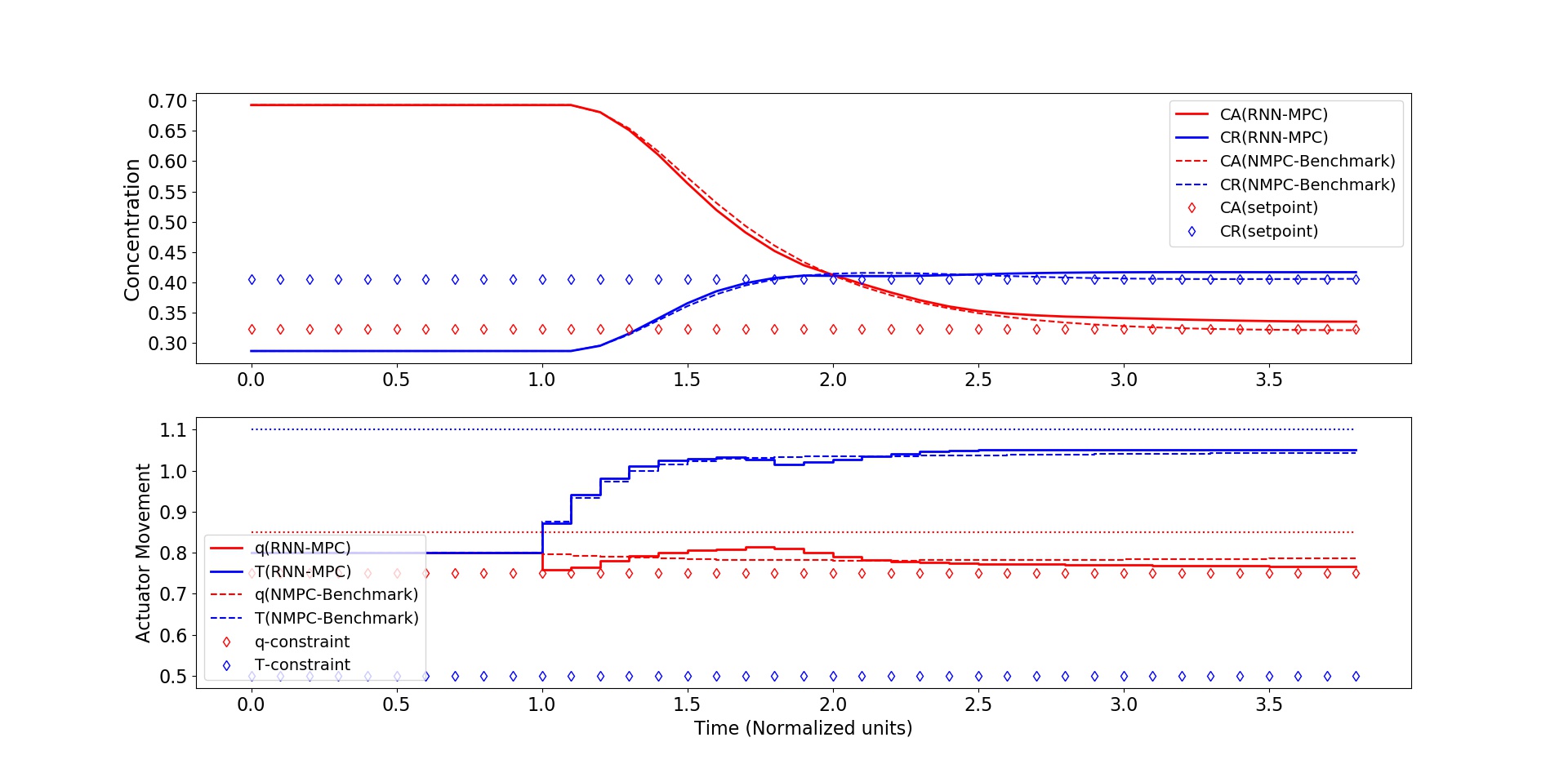} 
\caption{2000 nodes; 2 RNN layers - startup}
\label{fig:N2000_startup}
\end{subfigure}
\begin{subfigure}{1.00\textwidth}
\includegraphics[width=\linewidth]{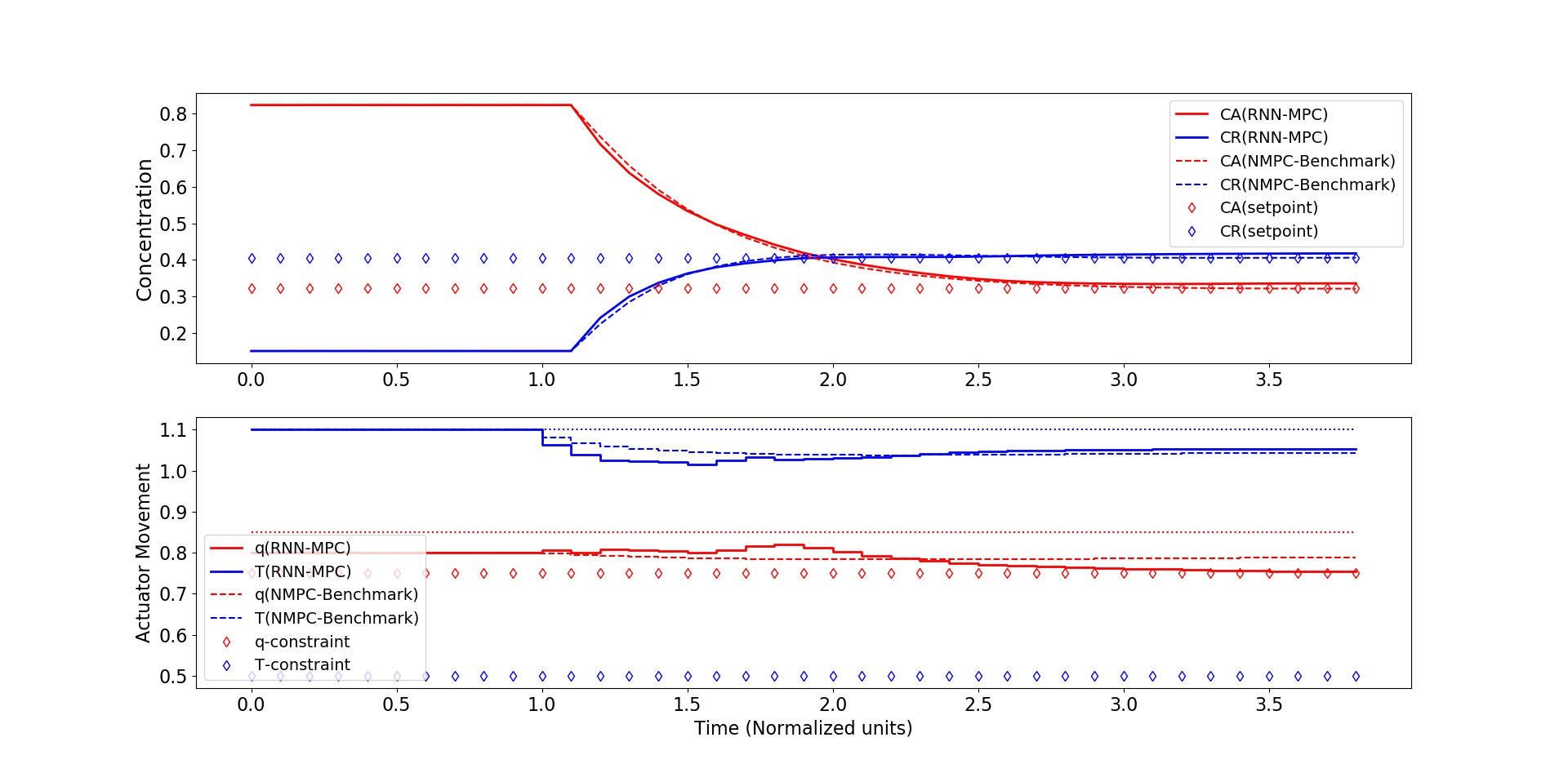}
\caption{2000 nodes; 2 RNN layers - recovery}
\label{fig:N2000_recovery}
\end{subfigure}
\caption{Closed-loop performance with 2000 nodes, 2 layers}
\label{fig:cl_N2000}
\end{figure}
\end{document}